\documentclass[usenatbib]{mn2e}
\usepackage{psfig}

\def\gsim{\mathrel{\raise0.35ex\hbox{$\scriptstyle >$}\kern-0.6em
\lower0.40ex\hbox{{$\scriptstyle \sim$}}}}
\def\lsim{\mathrel{\raise0.35ex\hbox{$\scriptstyle <$}\kern-0.6em
\lower0.40ex\hbox{{$\scriptstyle \sim$}}}}
\def\gs{\mathrel{\raise0.35ex\hbox{$\scriptstyle >$}\kern-0.6em
\lower0.40ex\hbox{{$\scriptstyle \sim$}}}}
\def\ls{\mathrel{\raise0.35ex\hbox{$\scriptstyle <$}\kern-0.6em
\lower0.40ex\hbox{{$\scriptstyle \sim$}}}}

\def\kms{\,\hbox{km}\,\hbox{s}^{-1}}
\def\Msol{\mathrel{\rm M_{\odot}}}
\def\Wm2{\,\hbox{W}\,\hbox{m}^{-2}}
\def\apj{ApJ}
\def\apjl{ApJL}
\def\mnras{MNRAS}

\def\aap{A\&AP}
\def\aj{AJ}

\def\nat{Nature}

\begin{document}

\title[Mass Decomposition of z=1 Giant Luminous Arcs]{The Baryonic and
  Dark Matter Properties of High Redshift Gravitationally Lensed Disk
  Galaxies}

\author[Salucci et al.]{
\parbox[h]{\textwidth}{
P.\ Salucci$^{1}$,
A.\ M.\ Swinbank$^{2}$,
A. Lapi $^{1}$,
I. Yegorova$^{1}$,\\
R.\ G.\ Bower$^{2}$,
Ian Smail$^{2}$,
G.\ P.\ Smith$^{3}$}
\vspace*{6pt} \\
$^1$Astrophysics Sector, SISSA/ISAS, Via Beirut 2-4, I-34014 Trieste, Italy\\
$^2$ ICC, Department of Physics, Durham University, Durham\\
$^3$ School of Physics and Astronomy, University of Birmingham, Edgbaston, Birmingham, B15 2TT\\
$^*$Email: salucci@sissa.it\\
}
\setcounter{footnote}{0}

\maketitle

 \begin{abstract}
   We present a detailed study of the structural properties of four
   gravitationally lensed disk galaxies at $z=1$.  Modelling the
   rotation curves on sub-kpc scales we derive the values for the disk
   mass, the reference dark matter density and core radius, and the
   angular momentum per unit mass.  The derived models suggest that the
   rotation curve profile and amplitude are best fit with a dark matter
   component similar to those of local spiral galaxies.  The stellar
   component also has a similar length scale, but with substantially
   smaller masses than similarly luminous disk galaxies in the local
   universe. Comparing the average dark matter density inside the
   optical radius we find that the disk galaxies at $z=1$ have larger
   densities (by up to a factor of $\sim 7$) than similar disk galaxies
   in the local Universe.  Furthermore, the angular momentum per unit
   mass versus reference velocity is well matched to the local
   relation, suggesting that the angular momentum of the disk remains
   constant between high redshifts and the present day.  Though
   statistically limited, these observations point towards a spirals'
   formation scenario in which stellar disks are slowly grown by the
   accretion of angular momentum conserving material.
\end{abstract}


\section{Introduction}

It has long been known that the kinematics of spiral galaxies do not
show Keplerian fall-off in their rotation curves, but rather imply the
presence of an invisible mass component in addition to the stellar and
gaseous disks \citep{Rubin80,Bosma81,Persic88}.  In the local Universe,
observing the distribution of star light and mapping the gaseous
component through H{\sc i} it has been possible to build up a picture
of the how the baryonic component of disk galaxies is distributed and
how this relates to the underlying dark matter component (e.g.
\citealt{Persic96}).  Tracing the evolution of galaxy mass from
high-redshift up to the present day is only truly reliable if we can
observe the same components at early times. A pioneering study was
performed by \citealt{Vogt96} where a handful of rotation curves (RC's)
of objects at $z\sim 0.5$ was obtained using traditional longslit
spectroscopy on the 10 meter Keck Telescope.  However, at higher
redshifts (e.g. z$\gsim$1) galaxies are much fainter and have smaller
angular disk scale lengths than galaxies observed at low redshift,
therefore obtaining the spatial information required for detailed
studies is beyond the limits of current technology.  Indeed, mapping
the internal properties and dynamics of both the stellar and gaseous
components of galaxies at high redshift is one of the main science
drivers for the next generation of ground and spaced based telescopes
at many wavelengths (e.g. ELT, NGST, ALMA).

One way to overcome this problem is to use the natural amplification
caused by gravitational lensing to boost the size and flux of distant
galaxies which serendipitously lie behind massive galaxy clusters.
This technique is extremely useful since we are able to target galaxies
which would otherwise be too small and faint to ensure a sufficiently
high signal-to-noise spectroscopy in conventional observations.  As
such, gravitational lensing has been extensively used to make detailed
studies of distant galaxies: for a galaxy at $z$=1 with an
amplification factor ten, an angular scale of 0.6$''$ can correspond to
$\lsim$0.5\,kpc, sufficient to map also a small spiral.

The benefits of gravitational lensing are complemented by Integral
Field Spectroscopy (which produces a contiguous {\it two dimensional
  velocity} map at each point in the target galaxy).  This allows a
clean decoupling of the spatial and spectral information, thus
eliminating the problems arising from their mixing in traditional
long-slit observations.  It is therefore much easier to identify and
study galaxies with regular (bi-symmetric) velocity fields.

In this paper, we present a detailed study of four rotation curves
extracted from disk galaxies which have been observed through the cores
of massive galaxy cluster lenses.  These targets are taken from the
recent work of \citet{Swinbank03,Swinbank06a}.  They were observed with
the Gemini-North Multi-Object Spectrograph Integral Field Unit (GMOS
IFU)\footnotemark.  We concentrate on the galaxy dynamics as traced by
the [O{\sc ii}]$\lambda\lambda$3726.1,3728.8\AA\ emission line doublet.
The IFU data provide a map of the galaxy's velocity field in sky
co-ordinates.  To interpret this field the magnification and distortion
caused by the gravitational lensing effect is removed using detailed
models of the cluster lenses (see \citet{Smith05}, \citet{Kneib96},
\citet{Smith02Th} for details).  The primary constraints on the lens
models are the positions and redshifts of spectroscopically confirmed
gravitational arcs in each cluster.  The source plane velocity fields
of four systems which display regular (bi-symmetric) rotational
velocity fields i.e.\ resembling rotating disks, were reduced to
one-dimensional rotation curves from which the asymptotic terminal
velocity was extracted and compared with the galaxy luminosity
\citep{Swinbank03,Swinbank06a}.  The key advantage of using
gravitational lensing to boost the images of distant galaxies is that
we are less biases towards the most luminous galaxies.  Whilst
observational information on the distribution of the H{\sc i} disks in
galaxies at these early times would be wellcome, such observations will
have to wait for future instrumentation (e.g.\, ALMA).

\footnotetext{Programme ID: GN-2003A-Q-3. The GMOS observations are
  based on observations obtained at the Gemini Observatory, which is
  operated by the Association of Universities for Research in
  Astronomy, Inc., under a cooperative agreement with the NSF on behalf
  of the Gemini partnership: the National Science Foundation (United
  States), the Particle Physics and Astronomy Research Council (United
  Kingdom), the National Research Council (Canada), CONICYT (Chile),
  the Australian Research Council (Australia), CNPq (Brazil) and
  CONICET (Argentina).}

In this paper we use nebular emission lines to probe the kinematic of
the galaxies.  We extract one-dimensional rotation curves from the
velocity fields to infer the distribution of stellar and dark matter
components.  Finally, we compare our results with similarly luminous
disk galaxies in the local Universe.  Through-out this paper we use a
cosmology with $H_{0}=72\kms$, $\Omega_{0}=0.3$ and $\Lambda_{0}=0.7$,
$t_0=13.7 \ Gyr$.

\section{Data }
\label{sec:analysis}

\subsection{Sample Selection}

Our sample comes from observations of six gravitational arcs in
\cite{Swinbank06a}.  In order to avoid possible biases, the targets
were selected only to be representative of galaxies in the distant
Universe and no attempt was made to select galaxies with relaxed
late-type morphology.  We did, however, require that arcs were resolved
in both spatial dimensions so that a two dimensional velocity field
could be extracted from the IFU data.  This restricted our selection to
galaxies with moderate magnification.  From the sample of six galaxies,
four galaxies appear to have (relaxed) bi-symmetric velocity fields
with late-type morphologies and colours.  The rotation curves from
these four galaxies appear regular and we therefore restrict our
analysis to these arcs.  We stress that the    morphology, colours and
velocity fields of the four galaxies in this sample all strongly
suggest these galaxies are consistent with late type spirals (see
\cite{Swinbank06a}).

\subsection{Photometry}

From our optical/near-infrared imaging, we constrain the spectral
energy distribution (SED) of each galaxy.  Since the arcs usually lie
with a few arcs-seconds of nearby bright cluster galaxies, we calculate
the magnitude of the arcs in various pass-bands by masking the arc and
interpolating the light from the nearby cluster members.  The
background light is then removed and surface photometry in different
bands is obtained.

Using the cluster mass models the arcs are reconstructed to the
source-plane and the geometry and disk-scale parameters of the disks
are measured.  This is achieved by fitting ellipses to an isophote
of the galaxy image using a modified version of the {\sc idl
gauss2dfit} routine which fits an exponential profile to the
two-dimensional light distribution.  From this,  the
ellipticity,  the inclination and luminosities    and the disk scale lengths are obtained (see Table 2 and 3 in
\cite{Swinbank06a}).  These latter  are also   reported  below in Table 1.

\begin{figure*}
\centerline{
  \psfig{file=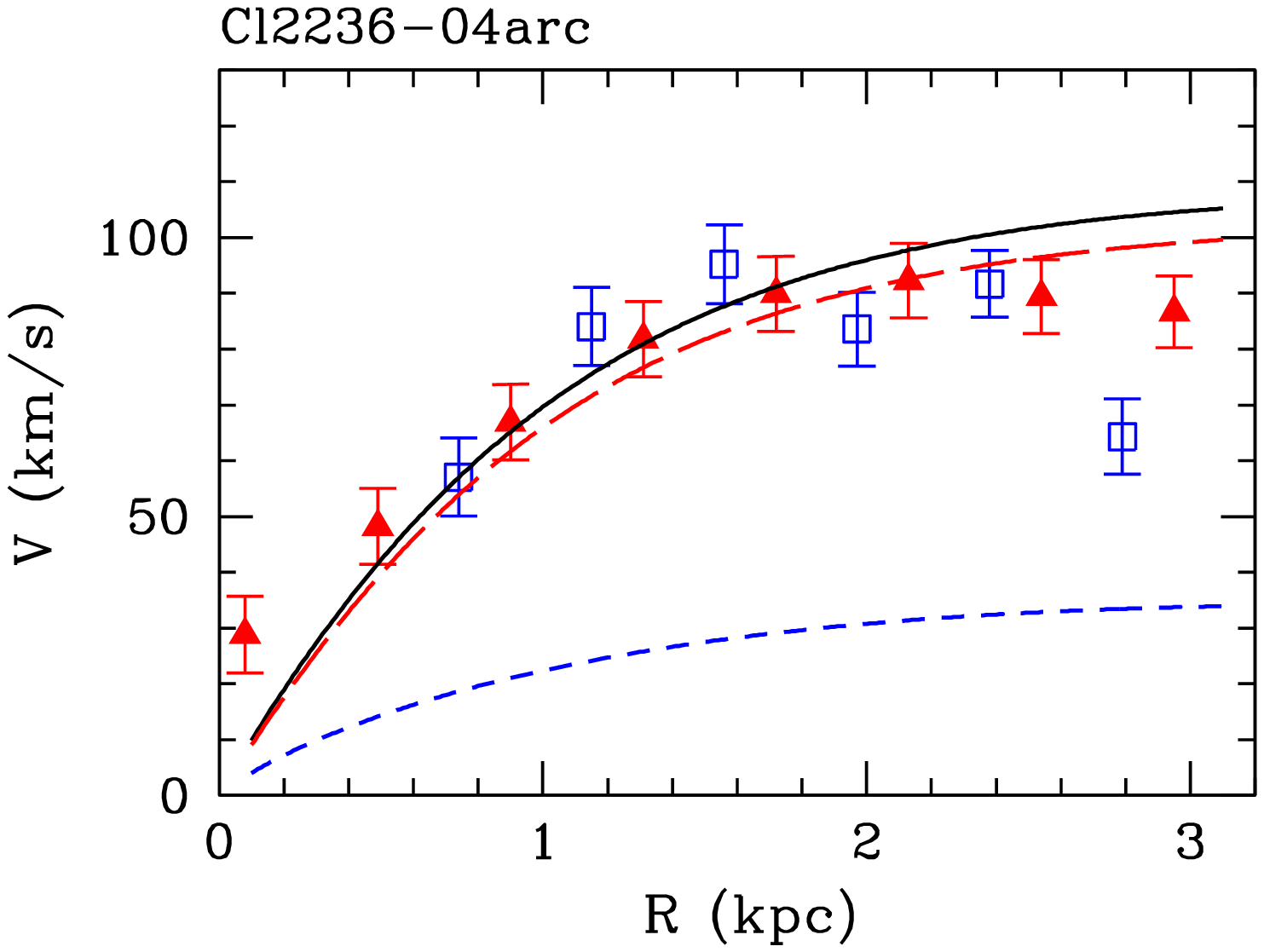,width=2.0in,angle=0}
  \psfig{file=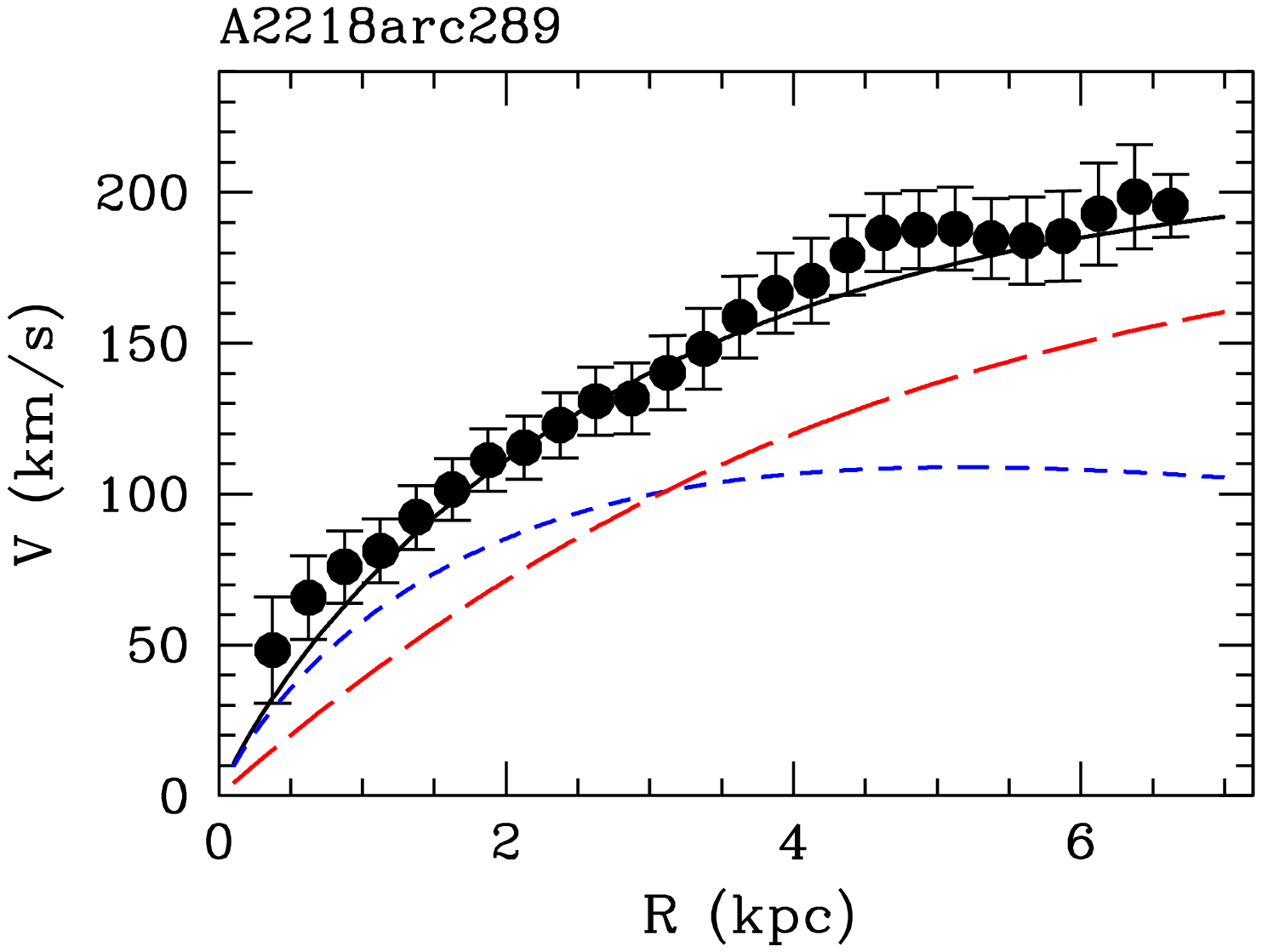,width=2.0in,angle=0}
}
\centerline{
  \psfig{file=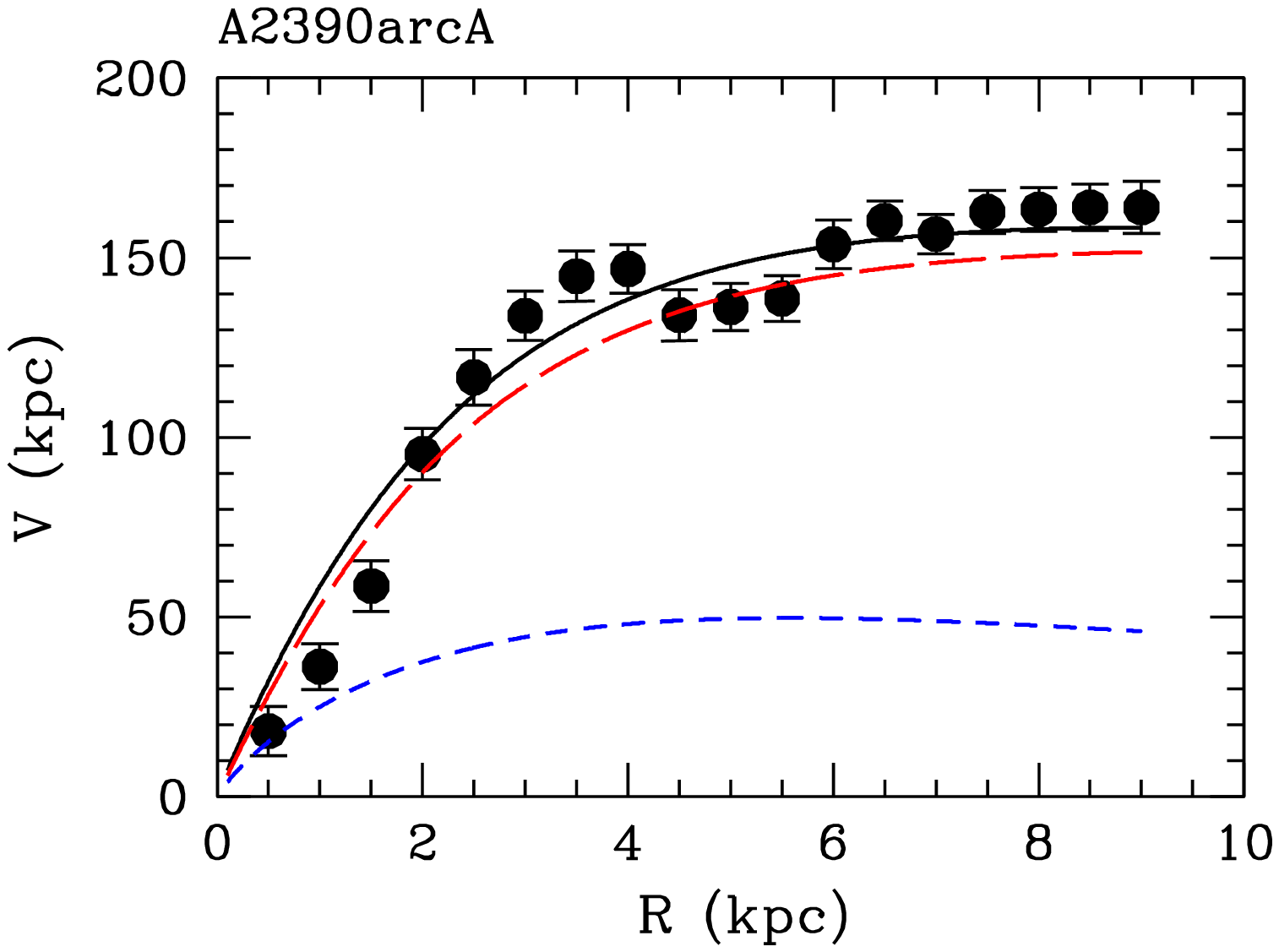,width=2.0in,angle=0}
  \psfig{file=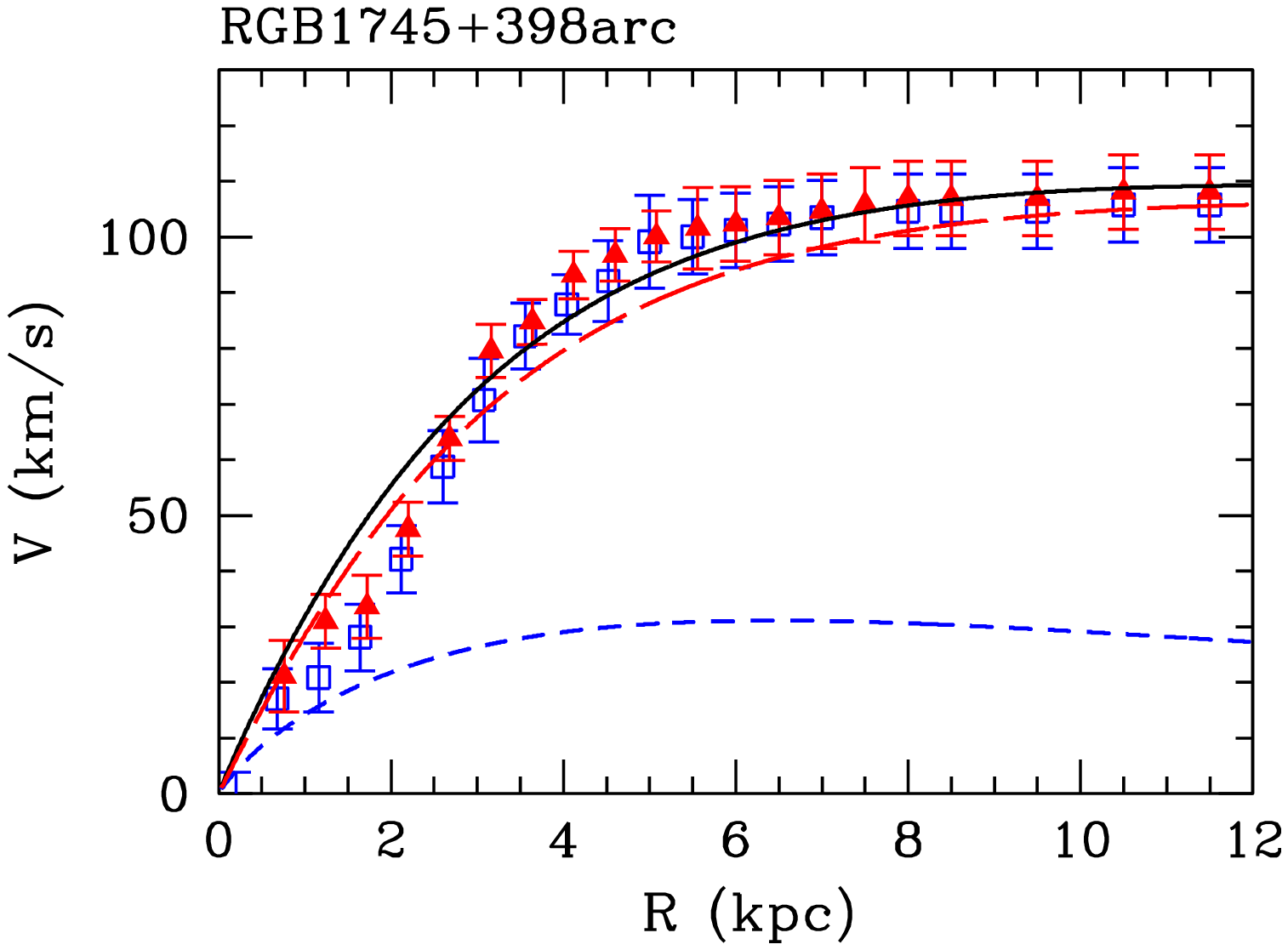,width=2.0in,angle=0}
}
\caption{  
  Filled circles represent the IFU rotation curves having enough data
  to allow the coaddition of the kinematics of the receding and the
  approaching arm, red and blue open circles represent the rotation
  curves along each arm in the other cases.  The dark-matter and
  stellar components are shown with long dash and short dash lines,
  whilst the model circular velocity is shown with a solid line.}
\label{fig:rot_curves}
\end{figure*}

\begin{table}
\begin{center}
\hspace*{-0.7cm}
{\scriptsize
\smallskip
{\sc Properties of the galaxies}
\hspace*{-0.7cm}
\begin{tabular}{lcccccccc}
\hline\hline
\noalign{\smallskip}
Arc             & $z$      & $M_{D}$            & $r_{0}$   & $\rho_0 $       & $R_{opt}$  & $v_{0h}$ \\
                &     & ($10^{10}\Msol$)  & ($kpc$)   & (10$^{-23}$g\,cm$^{-3}$) & $(kpc)$     & ($\kms$)   \\
\hline
\noalign{\smallskip}
A2390     & 0.912 & 0.40                   & 3.1      & 1.4        &  8.3       & 151      \\
RGB1745   & 1.056 & 0.18                   & 4.2      & 0.37       &  9.6       & 104      \\
A2218     & 1.034 & 1.70                   & 5.8      & 0.64       &  7.7       & 166      \\
Cl2236    & 1.116 & 0.12                   & 1.4      & 3.2        &  5.4       & 101      \\

\hline
\label{table:arcs}
\end{tabular}
}
\vspace{-0.5cm}
\caption{Derived structural parameters from the RC mass modelling.
  Error bars for $M_D$ are shown in \ref{fig:MD_rhom}, while the
  uncertainties in $r_0$, $v_{0h}$, $R_{opt}\equiv 3.2 R_D$ and $\rho_0$ amount to
  $30\%$, $15\%$, $15\%$, $40\%$ respectively.}
\end{center}
\end{table}

\subsection{One-Dimensional Rotation Curves}
\label{sec:onedRC}

In Fig.~\ref{fig:rot_curves} we show the one dimensional rotation
curves of the galaxies in our sample. These are extracted by sampling
the velocity field along the major axis cross section.  The zero-point
in the velocity is defined using the center of the galaxy in the
reconstructed source plane image.  The error bars for the velocities
are derived from the formal $3\sigma$ uncertainty in the velocity
arising from Gaussian profile fits to the [O{\sc ii}] emission in each
averaged pixel of the datacube.  For the mass modelling analysis we
folded the rotation curves on the kinematical center to ensure that
any small-scale kinematical sub-structure is removed.

\section{Modelling and Results}

The rotation curves of local spiral galaxies imply the presence of
an invisible mass component, in addition to the stellar and gaseous
disks. The paradigm is that the circular velocity field can be
characterized by:
$$V^2= V^2_{D} + V^2_{H} +V^2_{HI} \eqno(1)$$
where the subscripts denote the stellar disk, dark halo and gaseous
disk respectively.  From the photometry we model the stellar component
with a Freeman surface density \citep{Freeman70}:
$$\Sigma_D(R) = {M_D \over 2 \pi R_D^2}\, e^{-R/R_D} \eqno(2) $$
where $R_D$ is the disk lenght-scale, while $R_{opt} \equiv 3.2 R_D$ can
be taken as the "size" af the stellar disk, whose contribution to the
circular velocity is:
$$V^2_{D}(x) = {1\over 2}{GM_D\over {R_D}} (3.2 x)^2 (I_0K_0-I_1K_1) \eqno(3) $$
where $x=R/R_{opt}\equiv R/(3.2 R_D)$ and $I_n$ and $K_n$ are the
modified Bessel functions computed at $1.6~x$.  In small spirals the
H{\sc i} disk is an important baryonic component only for $R > 3 R_D$,
i.e. outside the region considered here. It is plausible that in
similar objects also at high redshifts, {\it inside} $3 R_D$, the H{\sc
  i} gas contributes to the gravitating baryonic mass by a very small
amount, residing at larger radii and it slowly infalls forming the
stellar disk.

For the dark matter component we take a spherical halo for which
$V_{H}^2 (R) = {G\, M_{H}(<R)/R}$.  Following the observational
scenario constructed in the local Universe \citep{Persic88,Salucci00}
we assume that it has the \citet{Burkert95} density profile (see also Salucci et al 2007):
$$\rho (R)={\rho_0\, r_0^3 \over (R+r_0)\,(R^2+r_0^2)}~, \eqno(4)$$
where $r_0$ is the core radius and $\rho_0$ the effective core density.
It follows that:
$$M_{H}(R) =   k \left[\ln\left( 1+\frac{R}{r_0} \right) -
  \tan^{-1}\left(\frac{R}{r_0} \right) +
  \frac{1}{2}\,\ln\left(1+\frac{R^2}{r_0^2}\right) \right]~\eqno(5) $$
with $k=6.4\,\rho_0\,r_0^3$ and of course $V^2_{H}(R)= G M_{H}(R)/R$
We note that the adopted velocity profile is a quite general: it allows
a distribution with a core of size $r_0$, converges to the NFW profile
at large distances and, for suitable values of $r_0$ can also mimic the
NFW or a isothermal profile, over the limited region of galaxy which is
mapped by the rotation curves.

\begin{figure*}
\centerline{
  \psfig{file=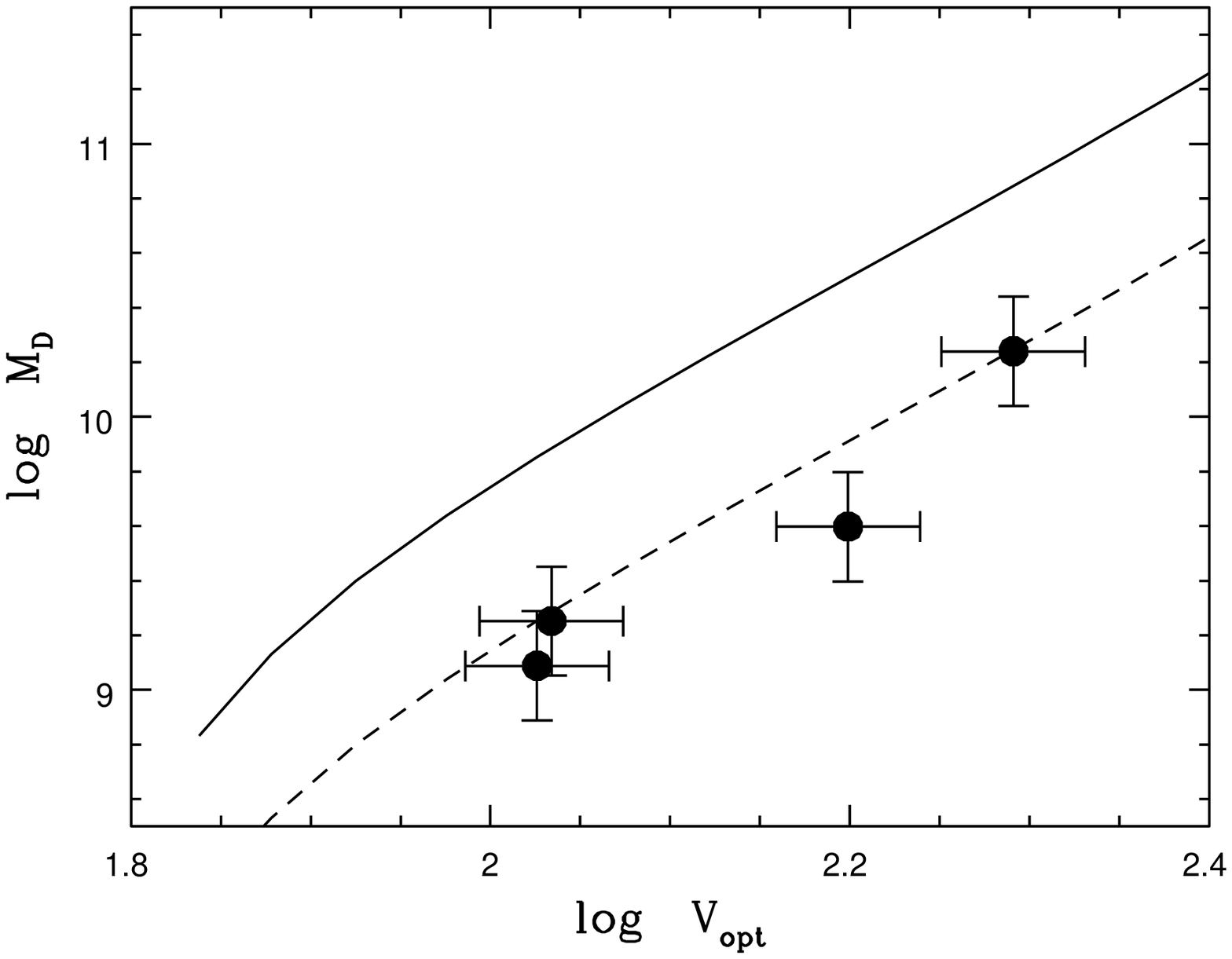,width=2.0in,angle=0}
  \psfig{file=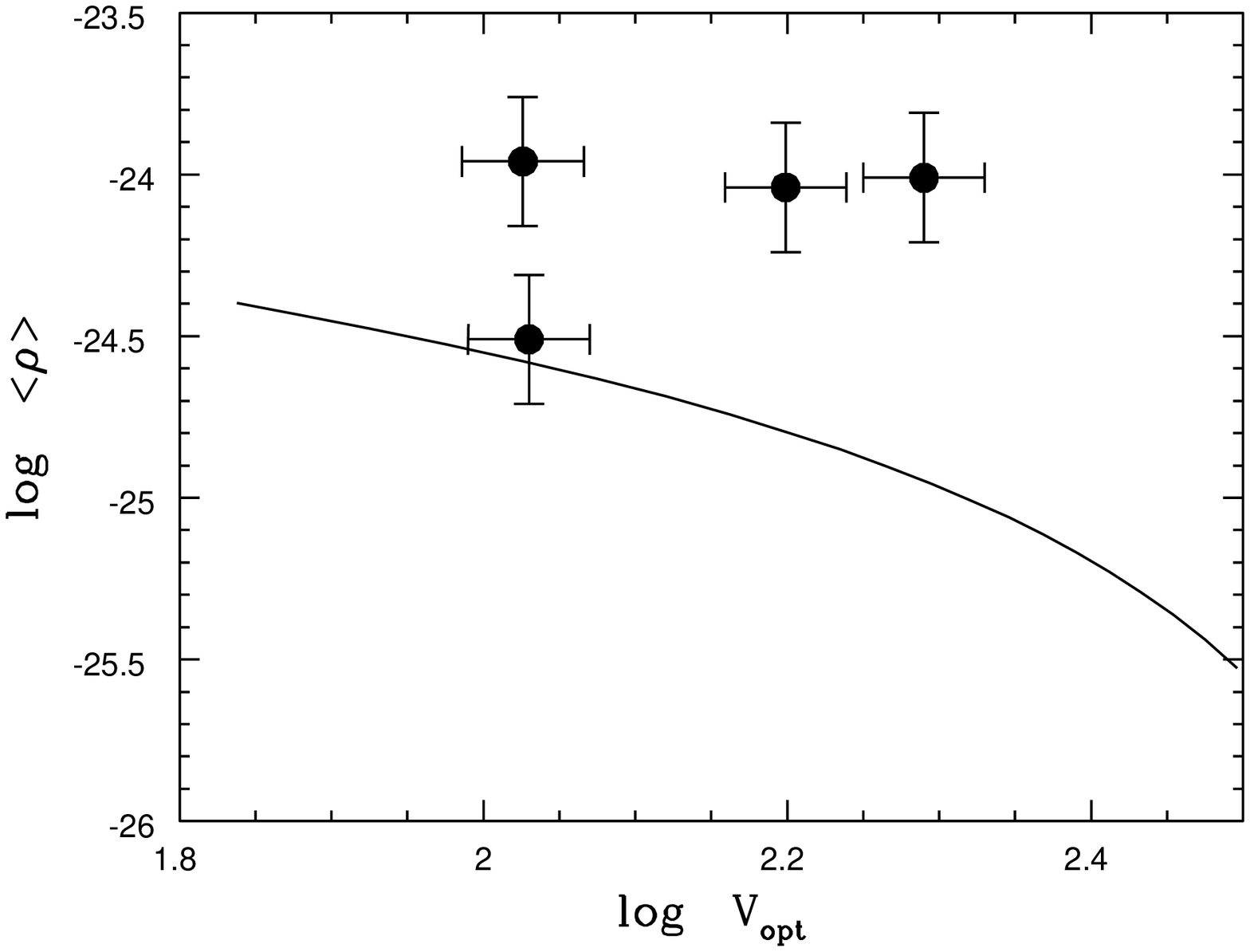,width=2.0in,angle=0}
  \psfig{file=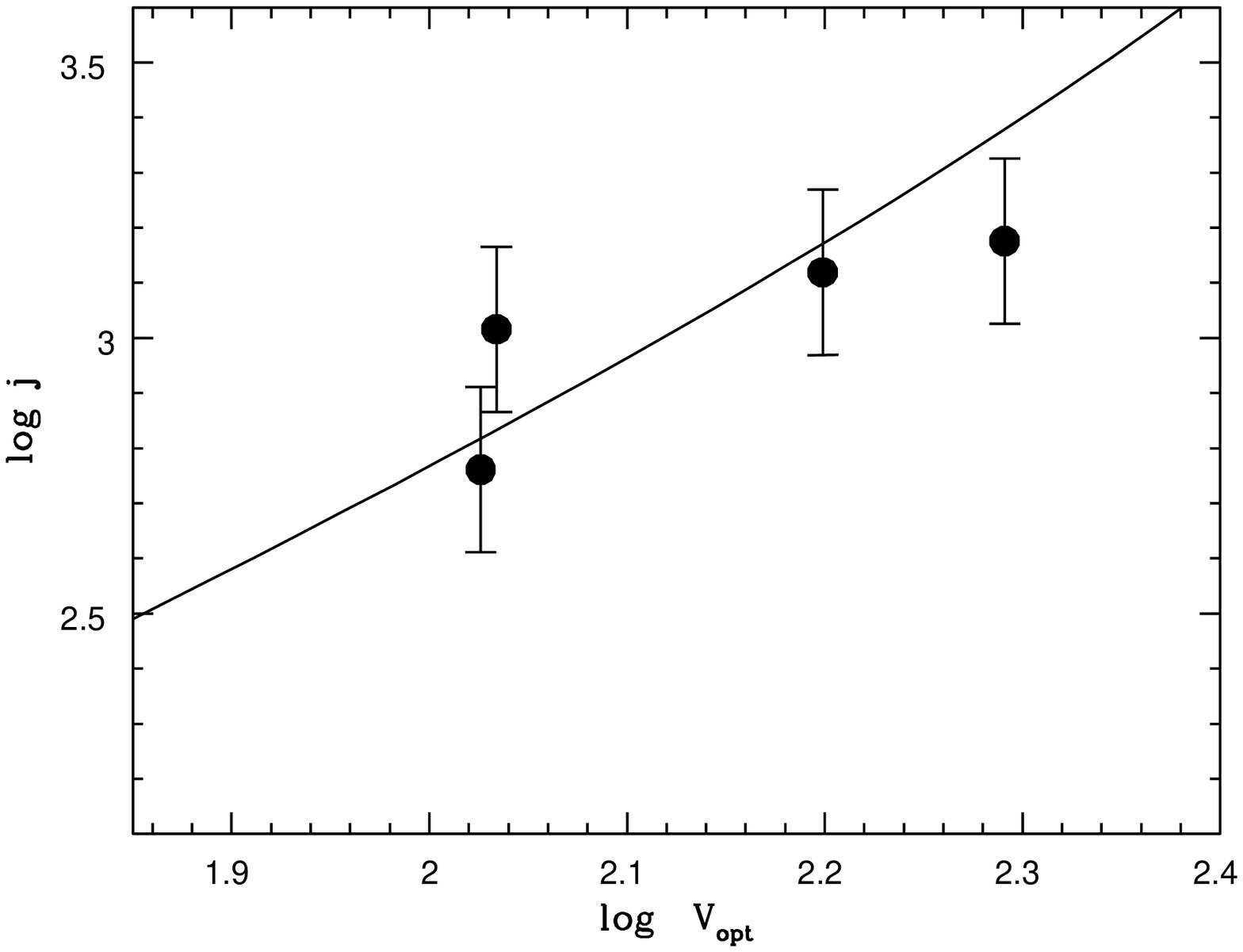,width=2.0in,angle=0}
}
\caption{Left: disk mass (in solar units) versus reference velocity (in
  $km/s$) compared with the $z$=0 relation ({\it solid} line) and with
  this scaled down by 0.6 dex ({\it dashed} line).  Middle: average
  dark matter density (in $g\ cm^{-3}$) versus reference velocity
  compared with the $z$=0 relation ({\it solid} line).  Right: disk
  angular momentum per unit mass $j$ (in $ km/s\,kpc$ units) versus
  disk mass compared with the $z$=0 relation ({\it solid} line). The
  uncertainties on the local relations are: 0.15 dex in $j$, 0.2 dex in
  $\log M_D$ and 0.3 dex in $\log <\rho>$. }
\label{fig:MD_rhom}
\end{figure*}

The mass model has three free parameters: the disk mass $M_D$, the core
radius $r_0$, and the central core density $\rho_0$.  The observations
extend out approximately $(2 - 3) R_D$, have 10-60 independent
measurements with an observational error of $3\%-10\%$ in their
amplitude, of 0.05-0.2 in their slopes $dlog V/dlog R$.  
The error   in the estimate of  the disk inclination angles is negligible with respect to  the above.
These errors are
(understandably) higher than those associated with the best quality
local RC's and make it difficult to constrain the halo density profile
$\rho(R)$.  However, they are sufficiently small to yield a reliable
value for the halo mass, the average density inside a reference radius,
(which we chose to be $ R_2 \equiv 2 R_D $ so that  $<\rho> \equiv
M_H(R_2)/(4/3 \pi R_2^3)$ and   the disk mass and a reasonable estimate of
the "core radius".

By reproducing the observed rotation curves with the models given by
equations 1-5 we derive the best fit parameters for each galaxy and
overlay the resulting mass model onto each rotation curve in
Fig.~\ref{fig:rot_curves}.  In Table 1 we report the main structural
parameters: the disk mass, the halo core radius, $\rho_0$, the optical
radius $R_{opt} $ and $ v_{0h}\equiv V_H(R_{opt})$, i.e.  the halo
contribution to circular velocity at the optical radius.
   
In all our rotation curves the {\it amplitude} and the {\it profile} of
the stellar disk contribution can not reproduce the observed RC rise
between 1.5 $R_D $ and the last measured radius.  This strongly
suggests evidence for the presence, at $z\approx 1$, of a dark matter
component of mass comparable to that found for local disk galaxies with
similar $V_{opt}$ (see Figures 2, 8 and 9 of Persic, Salucci \& Stel, 1996). We derive disk
masses ranging between $1 \times 10^9 M_\odot $ to $ 2 \times 10^{10}
M_\odot $ for galaxies with reference velocity $V_{opt}$ between
100$\kms$ and 200$\kms$.  These disk masses are smaller by a factor 2-4
than those of the local spirals with the same reference velocity which
are shown to follow Inner Baryon Dominance (see Salucci \& Persic, 1999).  The
high redshift stellar disks are sub-maximal disks.  Forcing a maximal
disk even in the "weak" implementation of Persic and Salucci (1990) leads to
unacceptable fits of our high-z RC's.
 
The best fit values for $r_0$ is of the order of $\sim 1.5 R_D$, which
is larger than usually compatible with a NFW profile, although the
error-bars on our data preclude any strong statement.

\begin{figure*}
\centerline{
  \psfig{file=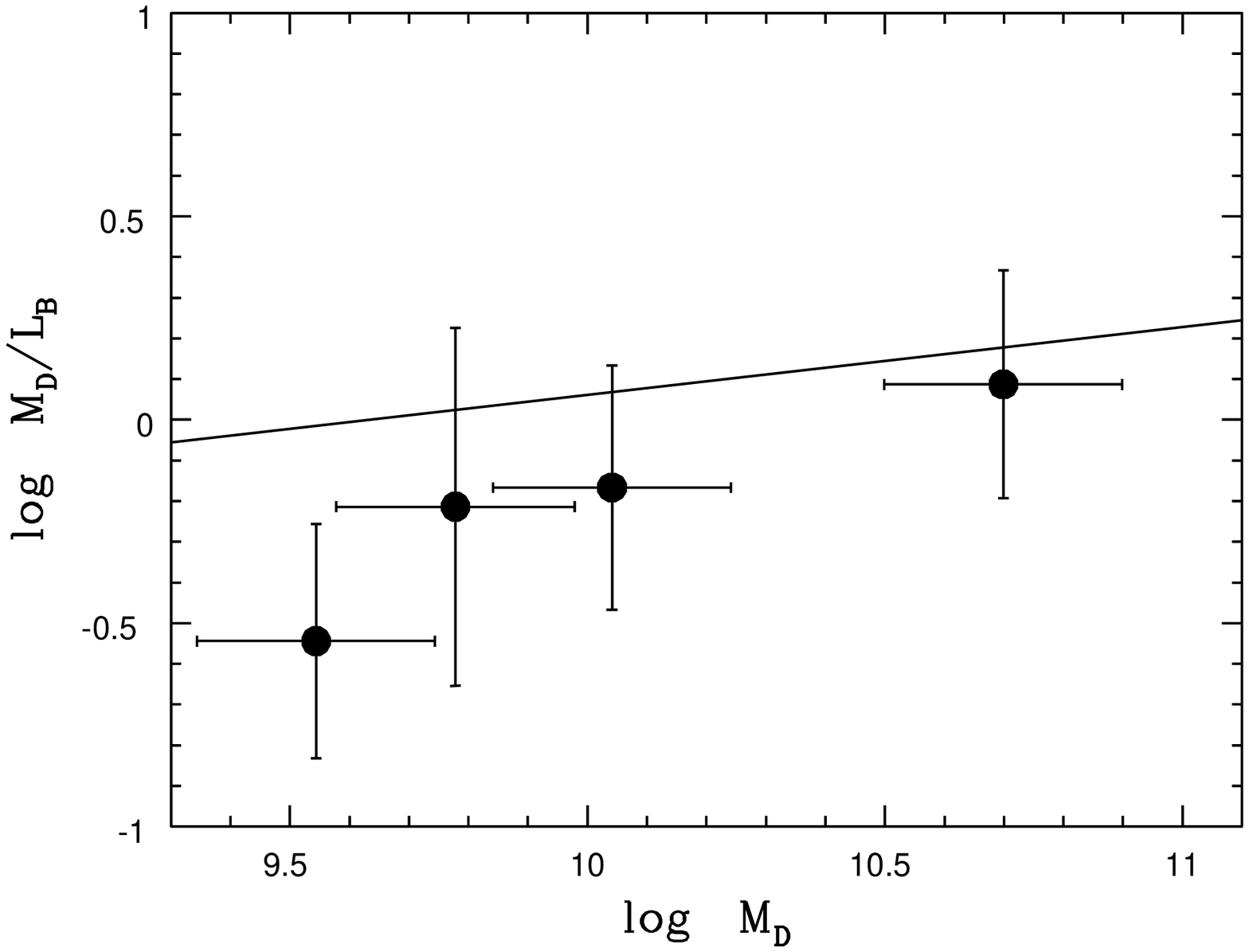,width=3.0in,height=2.65in,angle=0}
  \psfig{file=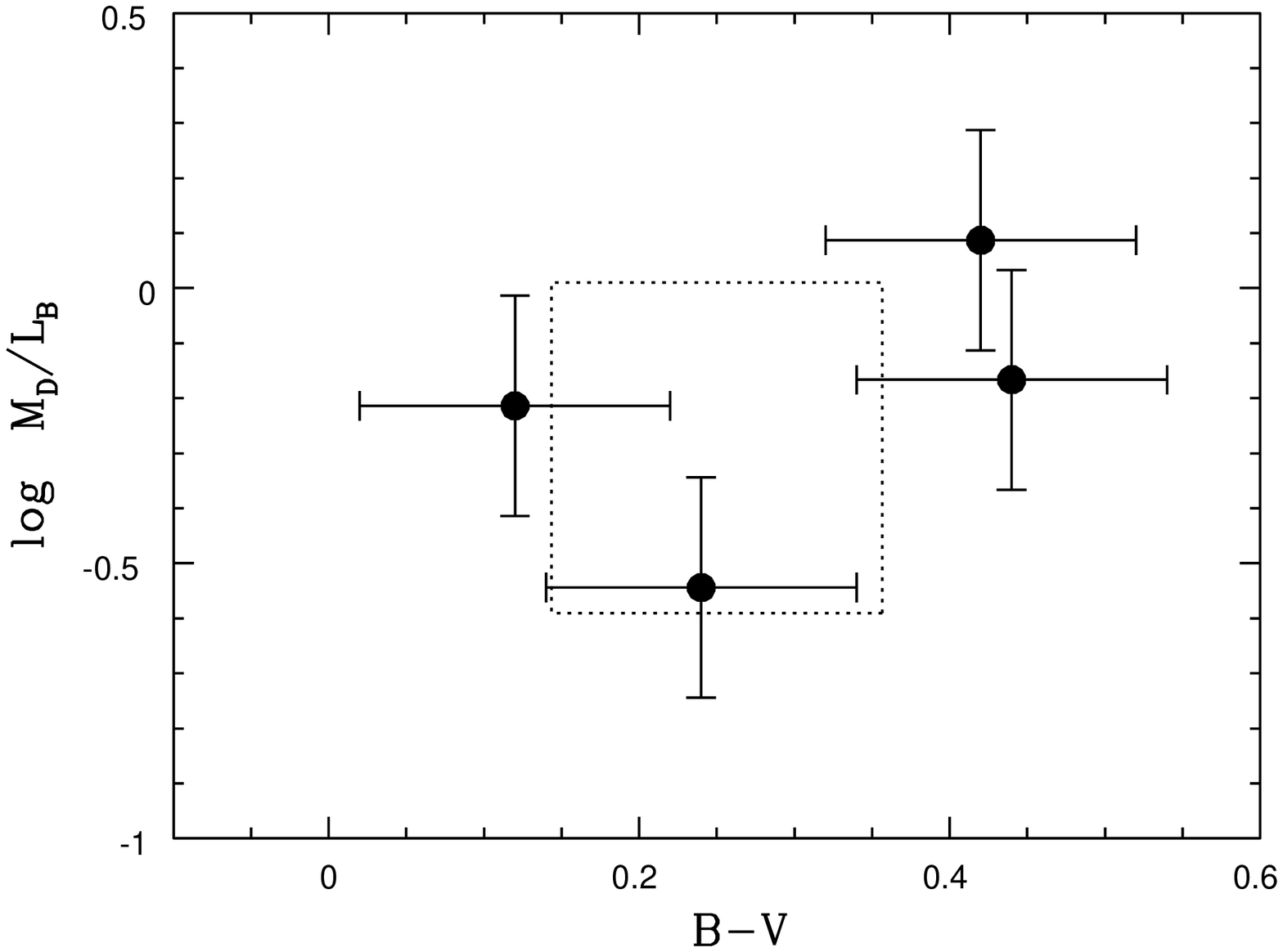,width=3.0in,height=2.65in,angle=0}}
\caption{{\it Left:} Mass-to-light ratio versus disk mass for the
  galaxies in our sample compared with the local disk galaxies from
  Shankar et al. (1996) (solid line).  The average offset corresponds
  to an excess in luminosity by a factor four.  {\it Right:}
  Mass-to-light ratio versus colors relations for our sample.  The
  square box denotes the predicted colours from a Bruzual \& Charlot
  single stellar population (see text for details).}
\label{fig:Mlp}
\end{figure*}

Since the gravitational lens model affects the source-plane
reconstruction of the galaxies, we must test for the effect that this
has on the resulting rotation curves.  We reconstruct the galaxies
using the family of lens models that inhabit the $\Delta \chi^2$ contour
corresponding to $1{\sigma}$ confidence interval relevant to each
cluster.  For example, the model of RGB\,1745 has five free parameters,
and the lens modeling uncertainties are therefore derived by
considering models within the ${\Delta}{\chi}^2{=}5.89$ contour. From
each reconstruction we extract the one-dimensional rotation curve and
apply the analysis outlined above and find the maximum variations are:
$\Delta\log \ M_{D}=0.03$, $\Delta\log<\rho>=0.02$,
$\Delta\log(V_{opt})\lsim5\kms$.  Thus the uncertainties in the
gravitational lens modelling is negligable compared to the
uncertainties in the RC mass modelling.

A cosmological significance of our result is evident in
Fig~\ref{fig:MD_rhom} where we compare the disk mass, the mean dark
matter density within the optical radius and the angular momentum per
unit mass, all as a function of reference velocity.  These are compared
to similar properties of the local objects \citep{Shankar06}.
Figure~\ref{fig:MD_rhom} strikingly shows that high redshift spirals,
modulo an offset of $0.6^{+0.1}_{-0.15}$ dex, are on the same $log M_D
$ versus $log V_{opt}$ relationship found for local spirals
  arising from the systematic structural properties of
their mass distribution \citep{Tonini}, see also \citep{Salucci93} .

In Fig 2,  from the values of  $<\rho>$,   a quantity that differently
from $\rho_0$ is weakly affected by the RC 1-$\sigma$ fitting uncertainties,  
 it is   apparent that 
   the DM halos of  $z=1$ disk galaxies  are denser by $0.7^{+0.1}_{-0.2}$ dex 
   than  those around similarly luminous $z$=0 spirals.  The evidence that spiral
disks at $z$=0 and $z$=1 have the same structural relationships is
further supported by observations of the evolution of the Tully-Fisher
relation (which correlates the disk mass with $V_{opt}$).  In our
and in other independent samples  (Swinbank et al 2006, , Vogt et al 1996, 
 Bamford et al 2005) the galaxies at $z=1$ show a TF relation with a
slope similar to that of the local TF, but with an offset compatible
with that found in  the present  work from  the disk mass {\it vs}  rotation
velocity.  This suggests that  from $z$=1 to $z$=0, the stellar disk masses
$M_{D}$ of a spiral   has grown by a factor
$\sim 4^{+1}_{-2}$, that leads to just a modest increase in the DM dominated quantity $V_{opt}$ .

Further evidence that $z=1$ disk galaxies are related to present day
spirals is provided by the relationship between angular momentum per
unit mass ($j$) versus the reference velocity, as shown in
Fig.~\ref{fig:MD_rhom}.  This well theoretically motivated relation
(e.g. \citep{Tonini})  can be considered as the imprint of the process of the
formation of disks inside dark matter halos  related to  the cosmological
properties of halo spin parameters.  As Figure~\ref{fig:MD_rhom} shows,
there appears to be no evolution in this crucial relationship between the the comological time at 
which we observe these spirals,    $z=1, t=6 \ Gyr$ and  the present time, 
 $z=0, t=13.7\  Gyr$.  This agreement is remarkable: it
establishes a link between local and high redshift disks, supporting
the idea that the angular momentum remains constant during the
evolution of a disk system from high redshift to the present day.
 
The kinematical estimate of the disk mass allows us to derive the
mass-to-light ratios for our disk systems as a function of luminosity
and colour.  In Fig.~3 we compare the disk mass and colour as a
function of mass-to-light ratio compared to the relation in local
spirals \citep{Shankar06}.  In order to compare directly with local
relations, we consider a simple passive evolution model for the
luminosity evolution.  For a single stellar population the zero-point
of the local relation is decreaed by a factor $log (13.7/6)$ which
accounts for the passive evolution of a {\it single} stellar population
from $z=1$ to $z=0$.  As Fig.~3 shows the mass-to-light ratios as a
function of galaxy ($B-V$) colour are in broad agreement with
predictions of a single stellar population which is $\sim1Gyr$ old
\citep{Bruzual03}, although clearly photometry at other wavelengths
(such as rest-frame K-band) would allow a more detailed decomposition
of the stellar popultations in these galxies.

\subsection{Cosmological Implications}

Figure~2 shows that at a radius corresponding to $V_{opt}$ the high
redshift galaxies are significantly denser than comparably luminous
local disk-galaxies: the average offset is about 0.6 dex in
$log(<\rho)>$.  Although we can not exclude that dynamical processes
occur between $z=1$ and $z=0$ to reduce the dark-matter density in the
luminous regions, this offset is naturally explained if the halos
embedding these disk galaxies formed at earlier times than the halos
around similarly massive $z=0$ spirals.  In this framework we estimate
the ratio between the virialization redshift of the local galaxies and
and that of the galaxies in our sample.  Since
$\rho_v\propto~\Delta(z_v) (1+z_v)^3$ where $\rho_v$ and $z_v$ are
average density and redshift at virialization and $\Delta_v$ is known,
for $z_{0}=1$, $z_{v}=1.7$, which corresponds to $t_v=6 \ Gyr$.

Assuming that our sample is a fair representation of disk galaxies at
$z\sim 1$ and that these are approximately coeval, from the comparison
of their structural properties with those of $z=0$ spirals, the
following simple picture emerges: a present day spiral, with a given
circular velocity, half-light radius and the angular momentum per unit
mass, at redshift $1$ had similar values for these quantities, but a
smaller stellar mass: $<M_\star (t_{obs})/ M_\star (t_0)> \simeq 0.3$.
This induces a scale for the average SFR in the past 8 Gyr: $ \sim ~
0.75 M_\star (t_{0}) /(t_{obs}-t_{0}) \sim \ 1 (M_\star(t_0)/ (10^{10}
M_\odot) M_\odot/yr$.  With these disks having an average age of 1\,Gyr
at $z=1$ we can also derive an "early times" average SFR $ \sim ~ 0.25
M_\star (t_{0}) / (1 Gyr) \sim 3 (M_\star(t_0)/ (10^{10} M_\odot)
M_\odot/yr$ which points towards a declining SFR history.

The marked increase of the luminosity per unit stellar mass in objects
at high redshifts with respect to their local counterparts has the
simplest explanation in a passive evolution of the starforming disks.
Obviously, this simple picture requires us to assume that the high
redshift systems are the direct counter-parts of similar rotation speed
spirals at low redshift.

\section{Discussion and Conclusions}
\label{sec:discussion}

In this study, we have investigated the detailed properties of four
disk galaxies at $z=1$.  These galaxies were observed at high
spatial resolution thanks to the boost in angular size provided by
gravitational lensing by foreground massive galaxy clusters and
allow a much more detailed comparison with local populations than
usually possible for galaxies at these early times.  Modelling the
one-dimensional rotation curves with those of \citet{Persic96} we
derive best fit parameters for the total dynamical mass, the core
radius, the effective core density and the angular momentum per unit
mass.

The best fit model rotation curves to the data show that the amplitude
and profile of the stellar disk componentcan not unambiguously
reproduce the rise in the rotation curve without a dark matter
component.  Comparing the average dark matter density inside the
optical radius we find that the disk galaxies at $z=1$ have larger
densities (by up to a factor of $\sim 7$) than similar disk galaxies in
the local Universe.  In comparison, we find that the angular momentum
per unit mass versus reference velocity is well matched to the local
relation suggesting that the angular momentum of the disk remains
constant between high redshift and the present day.  Though
statistically limited, these observations point towards a spirals'
formation scenario in which stellar disk are slowly grown by the
accretion of angular momentum conserving material.  Our result, also
consistent with the theoretical evolution of the angular momentum of
disks from semi-analytic models from $z$=1 to $z$=0 which show the
modest offset of $\Delta$j$\lsim$0.2 kpc.$\kms$ for objects with
circular velocities between 50 and 300$\kms$ (\citet{Cole2k,Bower06}).
These reults provide an evolutionary link between the disk systems we
observe at redshift $z\sim$1 and the present day population of spirals.

While these results are based on data of only four objects, they
nevertheless show the power which gravitational lensing can have on
studying the internal properties of high redshift galaxies.  In
particular, these observations should be viewed as pathfinder science
which will soon be routine with Adaptive Optics Integral Field Unit
observations on eight and ten meter telescopes (e.g.
\citealt{Genzel06}).  Moreover, by combining these results with
upcoming telescopes and instruments (e.g. ALMA) which will be sensitive
enough to map the H{\sc i} content of galaxies to $z=1$, the exact
relation between gas, stars and dark matter can be probed in much more
detail.

\section*{acknowledgments}
We would like to thanks the anonymous referee for  his/her   
suggestions which improved the content and
clarity of this paper.  We thank Carlton Baugh for providing the
theoretical evolution of disk mass in galaxies from {\sc galform}.  We
also thank Gigi Danese for useful discussions.  AMS acknowledges
support from a PPARC Fellowship, RGB acknowledges a PPARC Senior
Fellowship and IRS and GPS acknowledges support from the Royal Society.

\end{document}